\documentclass[aps,pra,twocolumn,unsortedaddress]{revtex4}
\usepackage{epsfig,pstricks,graphicx}
\usepackage{amsmath,amssymb}
\usepackage{color}

\def\DJo{$\;$\kern-.4em \hbox{D\kern-.8em\raise.15ex\hbox{--}\kern.35em okovi\'c}}
\def\CC{{\rm\kern.24em \vrule width.04em height1.46ex depth-.07ex
\kern-.30em C}}
\def\RR{{\rm
         \vrule width.04em height1.58ex depth-.0ex
         \kern-.04em R}}
\def\P{{\rm I\kern-.25em P}}
\def\id{{\rm 1\kern-.22em l}}

\def\up{\uparrow}
\def\down{\downarrow}
\def\trace{{\rm tr}\;}

\newcommand{\beq}{\begin{equation}}
\newcommand{\beqa}{\begin{eqnarray}}
\newcommand{\nbeqa}{\begin{eqnarray*}}
\newcommand{\eeq}{\end{equation}}
\newcommand{\eeqa}{\end{eqnarray}}
\newcommand{\neeqa}{\end{eqnarray*}}
\newcommand{\bra}[1]{\left\langle #1 \right |}
\newcommand{\ket}[1]{\left | #1 \right\rangle}
\newcommand{\braket}[2]{\left\langle #1 | #2 \right\rangle}

\newcommand{\bigfrac}[2]{\mbox {${\displaystyle \frac{ #1 }{ #2 }}$}}
\newenvironment{eqblock}[2]{\beq\label{#2}\begin{array}{#1}}{\end{array}
                                \eeq}
\newenvironment{neqblock}[1]{\[\begin{array}{#1}}{\end{array}\]}
\newcommand{\beqb}{\begin{eqblock}}
\newcommand{\eeqb}{\end{eqblock}} 
\newcommand{\nbeqb}{\begin{neqblock}}
\newcommand{\neeqb}{\end{neqblock}}

\begin{document}

\title{Possibility of generalized monogamy relations for 
multipartite entanglement beyond three qubits}
\author{Christopher Eltschka}
\affiliation{Institut f\"ur Theoretische Physik, 
         Universit\"at Regensburg, D-93040 Regensburg, Germany}
\author{Andreas Osterloh}
\affiliation{Institut f\"ur Theoretische Physik, 
         Leibniz Universit\"at Hannover, D-30167 Hannover, Germany}
\author{Jens Siewert}
\affiliation{Institut f\"ur Theoretische Physik, 
         Universit\"at Regensburg, D-93040 Regensburg, Germany}

\begin{abstract}
We discuss the possibility to interpret the residual entanglement
for more than three qubits in terms of distributed multipartite
entanglement, or, in other words, possible extensions of the 
Coffman-Kundu-Wootters monogamy equality to higher qubit numbers. 
Existing knowledge on entanglement in multipartite systems
puts narrow constraints on the form of such extensions. We
study various examples for families of pure four-qubit states
for which the characterization of three-qubit and four-qubit 
entanglement in terms of polynomial invariants is known.
These examples indicate that, although families with such extensions
do exist, a generalized monogamy equality cannot be found along those
lines.
\end{abstract}

\maketitle

\section{Introduction}
Getting insight into multipartite entanglement is one of the
challenges in quantum information theory. A seminal step torwards this
goal was the discovery of the analytic expression for pairwise 
qubit entanglement -- the concurrence of arbitrary two-qubit 
states~\cite{Hill97,Wootters98}. Interestingly, this
measure very soon lead to a further breakthrough as 
there is rather restricted freedom to distribute pairwise entanglement
in a three-qubit pure state. This constraint can be cast 
into the so-called {\em monogamy relation}~\cite{Coffman00}:
the total amount of entanglement for a given qubit 
(quantified by the {\em tangle}, or linear entropy) bounds
the sum
of two-qubit entanglement (measured by the two-tangle) of all pairs
with the qubit under consideration.

As for an arbitrary pure
three-qubit state, the discrepancy between tangle and the sum of two-tangles is non-zero
it was attributed to three-partite entanglement, the 
three-tangle~\cite{Coffman00}.  Interestingly it turned out that the three-tangle fulfills 
all requirements for an entanglement measure~\cite{MONOTONES,Duer00,VerstraeteDM03}
and therefore  it indeed quantifies the genuine three-party entanglement~\cite{Duer00}.
Later, Osborne and Verstraete presented a proof that also for arbitrary
pure $N$-qubit states the tangle
is a bound for the total amount of shared pairwise entanglement~\cite{Osborne06}.
However, even to date it is not clear whether also in the general case $N>3$ 
the difference between tangle
and the sum of two-tangles can be expressed in some way in terms
of quantities that quantify the distributed multipartite entanglement.

The first studies in this direction have been performed recently~\cite{Bai07,Bai08}
where specific pure four-qubit states have been analyzed with respect
to their tangle and concurrence. As a working hypothesis, the
authors took a straightforward extension of the monogamy relation
for granted. On that basis, they derived the mixed-state three-tangles and 
pure-state four-tangles 
as solutions of the resulting set of linear equations.
The conclusion from their analysis was that these three-tangles and four-tangles
are entanglement monotones only for some four-qubit pure states.
This result raises the question whether the underlying working hypothesis
is a good starting point at all.

In this work, we choose an alternative approach that is based on 
polynomial $SL(2,\CC)$ invariants as
multipartite entanglement measures for three qubits~\cite{Coffman00} 
and four qubits~\cite{Luque02,OS04,OS05,Levay06,Dokovic,DoOs08}. 
These invariants are entanglement monotones with respect
to stochastic local operations and classical communication (SLOCC)~\cite{VerstraeteDM03}.
The relevant ingredient is the analytical solution for the convex roof
of the three-tangle for rank-two mixed three-qubit states. 
A recent analysis provided solutions for various families of such states~\cite{LOSU,EOSU},
and even for rank-three states~\cite{Jung09}.
We mention that there are different approaches to describe monogamy properties 
of multipartite entanglement, e.g., in terms of different 
entanglement measures~\cite{Ou07,Bandyopadhyay07,Yu08,Sharma08,Kim09}, and
also for continuous-variable systems~\cite{Adesso07-1,Adesso07-2}.

It is important to note that monogamy relations emerge from the concept of distributing
entanglement in various ways among many parties (quantified by the corresponding measures)
and thus implicitly generate 
also a classification of multipartite entangled states. On the other hand, it
is not a priori clear which one among the many existing 
approaches to classify 
multipartite entanglement (e.g., Refs.~\cite{Duer00,VerstraeteDMV02,OS04,mandilara,Lamata07})
allows for complete generalization of monogamy.

In this paper, we first explain in detail which type of generalized monogamy relation
we would like to consider (Section II). In Section III and IV we present various examples
for states that do obey the specified type of monogamy, as well as counterexamples.
It turns out that there is a family of pure four-qubit states (which we call
``telescope states'') whose monogamy relation
relies on a straightforward extension of the three-qubit Coffman-Kundu-Wootters
equality. Conclusions are presented in Section V.

\section{Structure of generalized monogamy relations}

The fundamental quantities entering the
Coffman-Kundu-Wootters monogamy inequality 
for multipartite qubit systems are the tangle $\tau_1^{(j)}$ 
(or ``one-tangle'') 
of qubit number $j$
and the two-tangle $\tau_2^{(jk)}=C^2_{jk}$ of qubits number $j$ and $k$,
where $C_{jk}$ is the concurrence of qubits $j$ and $k$.
They are defined from the single and two-qubit reduced density matrices,
$\rho^{(1)}_j$ and $\rho^{(2)}_{jk}$,  of the $N$-qubit pure state 
$\rho=\ket{\psi}\!\bra{\psi}$ as
\beq\label{def-tau1}
\tau_1^{(j)} := 4\det \rho^{(1)}_j
\eeq
and
\beq\label{def-conc}
C_{jk} :=  \max\{0,2\lambda_{\rm max}-\trace \sqrt{R_{jk}}\}
\eeq
$\lambda^2_{\rm max}$ is the largest eigenvalue of the positive
Hermitean operator  
\beq
R_{jk} := \sqrt{\rho^{(2)}_{jk}}(\sigma_2\otimes\sigma_2)\rho_{jk}^*
                          (\sigma_2\otimes\sigma_2)\sqrt{\rho^{(2)}_{jk}} \ \ ,
\eeq
where $\sigma_\mu$, $\mu=1,2,3$ denote the Pauli matrices and $\sigma_0\equiv \id$.
In terms of these quantities the monogamy relation is expressed 
as~\cite{Coffman00,Osborne06}
\beq\label{CKW}
\mathcal{R}^{(j)}  :=  \tau_1^{(j)}- \sum_{k\neq j} \tau_2^{(jk)} \ge 0\ \ .
\eeq
For pure three qubit states, the residue $\mathcal{R}^{(j)}$  in Eq.~(\ref{CKW})
turns out to be an 
entanglement monotone, namely the three-tangle (or residual tangle): 
\beq
 \tau_1^{(j)}- \sum_{k\neq j} \tau_2^{(jk)} \ =\ \tau_3\ \ .
\label{CKW-eq}
\eeq
This is the celebrated Coffman-Kundu-Wootters monogamy equality~\cite{Coffman00}.
The three-tangle is most conveniently expressed as
\beqa
\tau_3(\psi)
 &=& \left| \bra{\psi^*}\sigma_\mu\otimes\sigma_2\otimes\sigma_2\ket{\psi}
\bra{\psi^*}\sigma^\mu\otimes\sigma_2\otimes\sigma_2\ket{\psi}\right|
\nonumber\\
&\equiv& \left| \bra{\psi^{\ast}}\bullet\bra{\psi^{\ast}} (\sigma_{\mu} \sigma_2\sigma_2)
         \bullet (\sigma^{\mu} \sigma_2\sigma_2)\ket{\psi}\bullet\ket{\psi} 
         \right|
\nonumber\\
&\equiv& \left| (\sigma_{\mu} \sigma_2\sigma_2)\bullet (\sigma^{\mu} \sigma_2\sigma_2)\right|
\label{def-tau3}
\eeqa
where \mbox{$\sigma_{\mu}\bullet \sigma^{\mu}=$}
                   $\sum_{\mu} G_{\mu}\sigma_{\mu}\bullet \sigma_{\mu}$
with $(G_0,G_1,G_2,G_3)=(-1,1,0,1)$.
That is, the three-tangle can be written as an expectation value of an antilinear
operator with respect to a two-fold copy of the state $\ket{\psi}$.
The ``$\bullet$'' 
in the second and third line of Eq.~(\ref{def-tau3}) represents a tensor product
and emphasizes the action of the operator on multiple copies~(see~\cite{OS04,OS05,DoOs08}).

The main question addressed in this article is whether, for arbitrary 
number of qubits $N$, the
residue in the monogamy relation~\eqref{CKW} can be expressed as
a sum of higher tangles,
i.e.\ polynomial $SL(2,\CC)$ invariants like the three-, four-,\dots ,
$N$-tangle. This question arises from the intuition of multipartite entanglement
as a resource that can be distributed in different ways among the parties~\cite{Coffman00}.

Let us first discuss the possible structure of such an extension in more detail.
An important restriction on the structure of a monogamy equality
arises from  the fact that, for an arbitrary qubit number, the 
inequality \eqref{CKW} saturates for W states~\cite{Coffman00},
i.e., $\sum_{k\neq j} \tau_2^{(jk)}/\tau_1^{(j)}=1$.
That is, the entanglement of these globally entangled multi-qubit states
is distributed in genuine two-qubit entanglement. 
Consequently, any generalized monogamy relation must be an additive extension to
the original monogamy equality, in which the one- and two-tangle must occur linearly 
and in the same way as in the original monogamy equality.

Thus, a generalized monogamy relation could be of the form
\beq
      \mathcal{R}^{(j)}=h(\tau_3^{(jkl)}, \tau_4^{(jklm)}, \ldots)
\label{arbmonog}
\eeq 
where $h$ is a positive function of the three-, four-, and higher
tangles involving the $j$th qubit. Note that the homogeneity degree
of the r.h.s.\ in Eq.~\eqref{arbmonog} has to be 4 as well. Keeping in
mind the conjectured character of multipartite entanglement as an 
additive resource, we restrict the r.h.s.\ in Eq.~\eqref{arbmonog}
to sums of the form
\[   h(\tau_3^{(jkl)}, \tau_4^{(jklm)}, \ldots) = \sum_{kl\neq j} f_3(\tau_3^{(jkl)})+
             \sum_{klm\neq j} f_4(\tau_4^{(jklm)})+\ldots
\]

A further restriction
comes from the fact that the three-tangle enters the monogamy equality
for pure three qubit states. This limits the tripartite 
entanglement monotone to coincide with the three-tangle on pure states.
A remaining freedom is to choose
the tripartite measure as the convex roof $\widehat{f(\tau_3)}$ of 
$f(\tau_3)$, where $f:[0,1]\to[0,1]$ is a strictly monotonous function,
and then to consider 
$f^{\!-1}(\widehat{f(\tau_3)})$
to obtain a homogeneous function of degree 4.
In the remainder of this article we consider monogamy relations for pure states
of at most four qubits, i.e., $N\le4$. Therefore, the only quantities involved
in the residue $\mathcal R$ are pure-state four-tangles and 
mixed-state three-tangles.
Hence, we analyze possible extended monogamy relations for four qubits
of the form
\beq
\mathcal{R}^{(j)} = \tau_1^{(j)}- \sum_{k\neq j} \tau_2^{(jk)} = 
\sum_{kl\neq j}f^{\!-1}(\widehat{f(\tau^{(jkl)}_3)})+\tau_4
\ \ .
\label{ourmonog}
\eeq
To this end, we will investigate various
families of interesting pure four-qubit states for which we are able
to compute the mixed-state three-tangle, and for which we can make 
statements about their genuine four-qubit entanglement.

It is worth mentioning that the residual tangle ${\cal R}^{(j)}$
vanishes not only for W states, but also for product states. 
This implies that $\tau_4=0$ for all product states, which
is a further justification to give major importance to multipartite 
entanglement measures with this property. The notion of
genuine multipartite entanglement measures as introduced in Refs.~\cite{OS04,OS05}
include the requirement for the measure to vanish on arbitrary product states.
Such measures form an ideal in the algebra of 
polynomial $SL(2,\CC)$ invariants~\cite{DoOs08}.

\section{An example}

In order to test the possibility of a generalized monogamy relation
in a simple but nontrivial case,
we may consider four-qubit states for which, however, the three-tangle
of the reduced density matrix has to be known.
Recently, the three-tangle of a whole family of mixed three-qubit
states has been found -- namely for rank-2 mixtures of GHZ states and W
states~\cite{LOSU,EOSU}. Therefore, we consider four-qubit states that
are purifications of those rank-2 states
\beqa\label{LOSU-states}
\ket{\Psi_p} & = & \sqrt{\frac{p}{2}}\left(\ket{1111}+\ket{1000}\right)\ +\ 
\nonumber\\
 &&  +\ \sqrt{\frac{1-p}{3}}\left( \ket{0100}+\ket{0010}+ \ket{0001}\right)\ \ .
\eeqa
In Refs.~\cite{OS04,OS05}, SLOCC invariants for genuine four-partite entanglement
in four-qubit states have been studied.
The four-tangle of the states (\ref{LOSU-states}) is measured only by 
the quantity
\beq
{\cal F}^{(4)}_1= \left|
                (\sigma_\mu\sigma_\nu\sigma_2\sigma_2)\bullet
                  (\sigma^\mu\sigma_2\sigma_\lambda\sigma_2)
        \bullet(\sigma_2\sigma^\nu\sigma^\lambda\sigma_2) \right|\; .
\eeq
The correct homogeneous degree $4$ is obtained via 
$\tau_4:=s\cdot \left({\cal F}^{(4)}_1\right)^{\frac{2}{3}}$. Note that 
the normalization of $\tau_4$ is not a priori clear. We account 
for it with a scaling factor $s$ and find 
\beq
\tau_4(\Psi_p)=s\cdot 4\sqrt[3]{\frac{2}{3}}p(1-p)\; .
\eeq
All other four-tangles are zero for this state.
Due to the permutation symmetry on the last three qubits,
there are two different values for the three-tangle:
$\tau_3^{(234)}$ has been determined in Ref.~\cite{LOSU}
and is zero for $p\leq p_0=\frac{4\sqrt[3]{2}}{3+4\sqrt[3]{2}}\sim 0.62$,
whereas from Ref.~\cite{EOSU} a direct calculation leads to
$\tau_3^{(123)}=\tau_3^{(124)}=\tau_3^{(134)}=0$ for all $p$.
Furthermore do all two-tangles including qubit number $1$ vanish
and all remaining two-tangles are equal and vanish for 
$p\geq p_c:=7-\sqrt{45}\sim 0.2918$~\cite{LOSU}.
The one-tangles are $\tau_1^{(1)}=4p(1-p)$ and
$\tau_1^{(j)}=\frac{(2+p)(4-p)}{9}$ for $j\neq 1$. 
The validity of a monogamy relation like Eq.~\eqref{ourmonog}
in the interval $0 \leq p\leq p_0$ would then imply
\beq
0=4p(1-p)-4s p(1-p)\ 
\eeq
and hence $s=1$ for the first qubit, and for the other qubits 
\beqa
0&=&\frac{(2+p)(4-p)}{9}-4s p(1-p)\ \ ;\ 
p_c\le p\le p_0
\\
0&=&\frac{3p(2-5p)}{9}+8(1-p)\sqrt{\frac{p(2+p)}{27}}- 4s p(1-p)\ 
\nonumber\\
&&
\mbox{for } 0\le p\le p_c\, .
\eeqa
No scaling factor $s$ can be found to adjust the monogamy relation
in all cases. We mention that the monogamy relations cannot even be satisfied on
average (that is, for the equally weighted sum of all 
one-tangles~\cite{Bai07}) with a $p$-independent $s$.
We conclude that no extended monogamy relation of the 
form~\eqref{ourmonog} can exist that includes the three-tangle and/or four-tangles,
and is valid for arbitrary pure four-qubit states.
An analogous analysis can be carried out for other families of states discussed
in Ref.~\cite{Bai07} and leads to the same conclusion (see Appendix).

\section{Telescope states}\label{Tel}

The findings in the previous section raise the question: 
are there any families of states for which monogamy persists?
A simple example is 
\beq\label{telescoped}
\ket{\Psi_{\rm tel}}:=\alpha\ket{1111}+\beta\ket{1000}+\gamma\ket{0110}\ \ .
\eeq
It is straightforward algebra to check that this state 
contains only two-tangle and three-tangle and that it satisfies
the monogamy relations of the form~\eqref{ourmonog} with $f\equiv \id$ 
for all four qubits. 
This specific state is an example for a pure quantum state in which
one (or more) single qubits
have a one-to-one correspondence to one (or more) single qubits of
a pure quantum state with a reduced number of qubits.
Such an $(N+m)$-qubit state emerges from a given pure $N$-qubit 
reference state by doubling one (or more) selected qubits
by what we will call {\em telescoping}.
This concept has been useful already in Refs.~\cite{OS05} 
for the creation of maximally entangled states for $q$ qubits 
from those known for $q-1$ qubits.
To give a specific example, 
from the three qubit reference state
$\ket{\cal M}=\sum_{k=0}^{1} m_k \ket{{\cal M}^{(k)}}_{12}\otimes\ket{k}_3$ 
the four-qubit telescoped state 
\beq\label{tel-state}
\ket{\cal T_M}=\sum_{k=0}^{1} m_k \ket{{\cal M}^{(k)}}_{12}\otimes\ket{kk}_{34}\; .
\eeq
is obtained by simply doubling the third qubit.
It is worth mentioning that the concept of telescoping is not reduced to this
specific form of extension. It is clear that instead of simple qubit doubling
\nbeqa
\ket{\psi}\otimes\ket{1}&\rightarrow& \ket{\psi}\otimes\ket{1}\otimes\ket{1}\\
\ket{\psi}\otimes\ket{0}&\rightarrow& \ket{\psi}\otimes\ket{0}\otimes\ket{0}\; ,
\neeqa
an arbitrary pair of orthonormal single qubit states, 
$\ket{\up}_{\vec{n}}$ and $\ket{\down}_{\vec{n}}$,  
can be used for the extension as
\nbeqa
\ket{\psi}\otimes\ket{1}&\rightarrow& \ket{\psi}\otimes\ket{1}
                      \otimes\ket{\up}_{\vec{n}}\\
\ket{\psi}\otimes\ket{0}&\rightarrow& \ket{\psi}\otimes\ket{0}
                      \otimes\ket{\down}_{\vec{n}}\; .
\neeqa
This amounts to a local unitary transformation on the added qubit {\em after}
telescoping.
Note that one can also apply a local unitary transformation on the original
state {\em before} telescoping, or even combine both.
It is interesting that telescoped product states are product states on
the partition induced by the telescoping procedure. Furthermore,
telescoping and qubit permutation do not commute.

In the following we analyze the 
entanglement pattern of the telescoped states.
After tracing out one of the telescoped qubits,
a biseparable density matrix is obtained. 
For the state \eqref{tel-state} this implies
\beq
\tau_3^{(123)}({\cal T_M})=\tau_3^{(124)}({\cal T_M})=0
\eeq
\beq
\tau_2^{(13)}({\cal T_M})=\tau_2^{(14)}({\cal T_M})=
\tau_2^{(23)}({\cal T_M})=\tau_2^{(24)}({\cal T_M})=0\ .
\eeq
Furthermore we have 
$\trace_{3,4} \ket{\cal T_M}\bra{\cal T_M}=
 \trace_{3} \ket{\cal M}\bra{\cal M}$ and therefore
$\tau_2^{(12)}({\cal T_M})=\tau_2^{(12)}({\cal M})$. Consequently, all
single-qubit reduced density matrices and hence all one-tangles coincide
for both states.
Invoking the three-qubit monogamy relation for the reference state $\ket{\cal M}$ 
fixes the values for the four-tangles entering the monogamy relations for
the four-qubit telescoped state
\beqa\label{tel-mismatch1}
\tau_{4;1} &=& \tau_3({\cal M})+ \tau_2^{(13)}({\cal M}) - \tau_3^{(134)}({\cal T_M})\\
\tau_{4;2} &=& \tau_3({\cal M})+ \tau_2^{(23)}({\cal M}) - \tau_3^{(234)}({\cal T_M})\; .
\label{tel-mismatch2}
\eeqa
By using the notation $\tau_{4;j}$ we allow for the possibility
that the monogamy relations on different qubits might be satisfied
mathematically with different four-tangles -- although, from a physical
point of view, this would be questionable.

The most surprising feature is  the  connection
between a certain two-tangle of the reference state
and a three-tangle of the telescope state.
To see this, consider the two decomposition states 
$\ket{{\cal M}^{(k)}}_{12}=:\alpha^{(k)}_{ij}\ket{ij}$ ($k=0,1$)
of $\rho^{(2)}_{23}({\cal M})=\trace_1 \ket{\cal M}\bra{\cal M}$ and
$\ket{{\cal T_M}^{(k)}}_{123}
=:\alpha^{(k)}_{ij}\ket{ijj}$ of 
$\rho^{(3)}_{234}({\cal T_M})=
\trace_1 \ket{\cal T_M}\bra{\cal T_M}$ where $i,j=0,1$  represent
the computational basis for the respective qubit (we drop the symbol $\sum_{ij}$
for brevity).
It is clear that any decomposition of $\rho^{(2)}_{23}({\cal M})$
is telescoped into a decomposition of 
$\rho^{(3)}_{234}({\cal T_M})$ and vice versa.
We now use the expression of the two- and three-tangle in terms of
antilinear expectation values~\cite{OS04} and obtain
\begin{widetext}
\beqa
\tau_2(\alpha^{(k)}_{ij}\ket{ij}) &=&
 \left|
 \alpha^{(k)}_{ij}\alpha^{(k)}_{lm}\alpha^{(k)}_{np}\alpha^{(k)}_{qr}
\bra{ij}\sigma_2\sigma_2\ket{lm} 
   \bra{np}\sigma_2\sigma_2\ket{qr}
            \right|
  \\
 \tau_3(\alpha^{(k)}_{ij}\ket{ijj}) 
&=&
 \left|
 \alpha^{(k)}_{ij}\alpha^{(k)}_{lm}\alpha^{(k)}_{np}\alpha^{(k)}_{qr}  
     \bra{ijj}\sigma_2\sigma_2\sigma_\mu\ket{lmm} \nonumber
   \bra{npp}\sigma_2\sigma_2\sigma^\mu\ket{qrr}
   \right|
  \\
&=& \left|
   \alpha^{(k)}_{ij}\alpha^{(k)}_{lm}\alpha^{(k)}_{np}\alpha^{(k)}_{qr} 
\bra{i}\bullet\bra{n}\sigma_2\bullet\sigma_2\ket{l}\bullet\ket{q}\!\!
\bra{j}\bullet\bra{p}\sigma_2\bullet\sigma_2\ket{m}\bullet\ket{r}\!\!
    \bra{j}\bullet\bra{p}\sigma_\mu\bullet\sigma^\mu\ket{m}\bullet\ket{r}
   \right|
\eeqa
\end{widetext}
Now it is sufficient to observe that for states of the computational basis
\beqa
&\bra{j}\bullet\bra{p}\sigma_2\bullet\sigma_2\ket{m}\bullet\ket{r}
\bra{j}\bullet\bra{p}\sigma_\mu\bullet\sigma^\mu\ket{m}\bullet\ket{r}&=
\nonumber\\
&=\bra{j}\bullet\bra{p}\sigma_2\bullet\sigma_2\ket{m}\bullet\ket{r}&
\eeqa
in order to establish that indeed 
\beqa
\tau_2^{(23)}({\cal M}) &=& \tau_3^{(234)}({\cal T_M}) \\
\tau_2^{(13)}({\cal M}) &=& \tau_3^{(134)}({\cal T_M}) \; .
\eeqa
Inserting these results into Eqs.~\eqref{tel-mismatch1}, \eqref{tel-mismatch2}
leads to
\beqa
\label{mismatch1}
\tau_{4;1}({\cal T_M}) &=& \tau_3({\cal M})\\
\label{mismatch2}
\tau_{4;2}({\cal T_M}) &=& \tau_3({\cal M})\; .
\eeqa
A simple calculation shows that all four-qubit $SL$-invariant tangles,
evaluated on telescope states \eqref{tel-state} contain the three-tangle of the 
reference state as a common factor. Hence,
if the reference state has no three-tangle, the telescope
four-qubit state has no four-tangle. Then the monogamy equality for
four qubits is readily satisfied on qubits $1$ and $2$.
Otherwise both four-tangles must coincide with the three-tangle of the reference state.

In order to analyze the general case where the four-tangle is non-zero,
we continue by verifying the monogamy relations for qubits $3$ and $4$.
We consider two cases:
{\em i)}\/ $\braket{{\cal M}^{(2)}}{{\cal M}^{(1)}}_{12} = 0$ and
{\em ii)}\/  $\braket{{\cal M}^{(2)}}{{\cal M}^{(1)}}_{12} \neq 0$.
In case {\em i)}\/ $\rho^{(2)}_{34}$ is separable and consequently 
$\tau_2^{(34)}({\cal T_M})=0$. In addition, the single qubit density matrices 
on sites $3$ and $4$ of the telescope state is identical to that on
site $3$ of the reference state. This implies
$\tau_1^{(3)}({\cal T_M})=\tau_1^{(4)}({\cal T_M})=\tau_1^{(3)}({\cal M})$
and we are ready to calculate the value of the four-tangle that appears in the monogamy relation
\beqa
\label{mismatch3}
\tau_{4;3}({\cal T_M}) &=& \tau_3({\cal M})\\
\label{mismatch4}
\tau_{4;4}({\cal T_M}) &=& \tau_3({\cal M})\; .
\eeqa
In case {\em ii)}\/ we can write uniquely
$\ket{{\cal M}^{(2)}}_{12}=\alpha \ket{{\cal M}^{(1)}}_{12} +\beta \ket{\perp}_{12}$
with $\braket{\perp}{{\cal M}^{(1)}}_{12}=0$, and a straightforward
calculation shows that the difference between the one-tangles
for reference and telescope state
compensate precisely with the resulting non-zero two-tangle 
$\tau_2^{(34)}({\cal T_M})=|\alpha m_0 m_1|^2$.
Therefore, Eqs~\eqref{mismatch3}, \eqref{mismatch4}
remain unaltered.
Summarizing the above discussion, we conclude that the
monogamy relation can be adjusted for telescope states
with a {\em single choice} for the value of the hypothetic four-tangle.

We will now use the trick involved in the equality of
the reference state two-tangle with the telescope state three-tangle 
in order to construct this unknown four-tangle. 
It can be derived from our finding that the monogamy equality
holds if and only if the four-tangle of the telescope state coincides
with the three-tangle of the reference state.
Using the identity 
\nbeqa
\tau_3(\psi)&=&\left|\bra{\psi^*}\sigma_\mu\sigma_2\sigma_2\ket{\psi}
             \bra{\psi^*}\sigma^\mu\sigma_2\sigma_2\ket{\psi}\right| \\
            &=&\left|\bra{\psi^*}\sigma_2\sigma_\mu\sigma_2\ket{\psi}
             \bra{\psi^*}\sigma_2\sigma^\mu\sigma_2\ket{\psi}\right| \\
            &=&\left|\bra{\psi^*}\sigma_2\sigma_2\sigma_\mu\ket{\psi}
             \bra{\psi^*}\sigma_2\sigma_2\sigma^\mu\ket{\psi}\right| 
\neeqa
we derive the relevant four-qubit polynomial $SL(2,\CC)$ invariants
(notations from Ref.~\cite{DoOs08}) as
\beqa
{\cal C}^{(4)}_{4;(1,4)} &:=& (\sigma_\mu\sigma_2\sigma_2\sigma_\nu)\bullet
                              (\sigma^\mu\sigma_2\sigma_2\sigma^\nu)\\
{\cal C}^{(4)}_{4;(2,4)} &:=& (\sigma_2\sigma_\mu\sigma_2\sigma_\nu)\bullet
                              (\sigma_2\sigma^\mu\sigma_2\sigma^\nu)\; .
\eeqa
Their absolute values give the corresponding four-tangles that fix all four monogamy
relations simultaneously.
Due to the relations\cite{DoOs08} 
${\cal C}^{(4)}_{4;(1,4)}={\cal C}^{(4)}_{4;(2,3)}$,
${\cal C}^{(4)}_{4;(1,3)}={\cal C}^{(4)}_{4;(2,4)}$,
${\cal C}^{(4)}_{4;(1,2)}={\cal C}^{(4)}_{4;(3,4)}$,
${\cal C}^{(4)}_{4;(1,4)}+{\cal C}^{(4)}_{4;(2,4)}+
{\cal C}^{(4)}_{4;(3,4)}=12 H^2$,
where $H=(\sigma_2^{\otimes 4})/2$ is the 4-concurrence from Ref.~\cite{Wong00},
we can also use $6 H^2-\frac{1}{2}{\cal C}^{(4)}_{4;(3,4)}$ as the four-tangle.
This implies that the three possible four-tangles 
${\cal C}^{(4)}_{4;(1,4)}$,
${\cal C}^{(4)}_{4;(2,4)}$, and  
$6 H^2-\frac{1}{2}{\cal C}^{(4)}_{4;(3,4)}$
have the same value on
telescope states generated by doubling qubit number $3$. 
It is clear that
doubling qubits $1$ or $2$ leads to analogous expressions.
Interestingly, the algebra of polynomial invariants of four-qubit telescope states
is generated by two independent elements only. When the third qubit is doubled, then
${\cal C}^{(4)}_{4;(1,2)}$ and ${\cal C}^{(4)}_{4;(1,3)}$ can be chosen as independent
generators. Consequently, all other four-tangles can be expressed uniquely as
a polynomial function of them.  

However, we stress that there is no {\em unique} genuine four-qubit entanglement 
measure that satisfies the four-qubit monogamy equalities for all four qubit 
telescope states.

The above-mentioned correspondence
of $q$-tangles of some $q$-qubit reference state to a set of $(q+m)$-tangles for
telescope states generated from the reference state 
by $m$-fold qubit doubling is a generic 
feature and appears for general $q$ and $m$.
Monogamy relations for $3+m$ qubits emerge directly from
the Coffman-Kundu-Wootters monogamy relation for pure three-qubit states.
The $(q+m)$-tangles satisfying the monogamy relations
are found  to depend on the specific qubit-doubling procedure that
creates the $(q+m)$-qubit state from its reference $q$-qubit state.
We conclude that no general extension to the monogamy relation~\eqref{CKW}
exist that includes $q$-tangles with $q> 3$, not even for telescope states.

An interesting representative of telescope states are the four-qubit cluster 
states~\cite{briegel01}
\begin{equation}
  \label{eq:2dcluster}
  a\ket{0000}-b\ket{0111}-c\ket{1100}+d\ket{1011}
\end{equation}
which have been considered in Ref.~\cite{Bai08} (up to a permutation of qubits $2$ and $3$).
As in Ref.~\cite{Bai07}, the authors in Ref.~\cite{Bai08}
take the existence of an extended monogamy relation for granted
and determine the three- and four-tangle in the state by solving
the resulting set of linear equations. 
We confirm the non-zero three-tangles to be
$\tau_3^{(124)}=4|ad-bc|^2$ and $\tau_3^{(234)}=4|ab-cd|^2$.
With the remaining one- and two-tangles the four-tangle that adjusts all 
four monogamy relations must take the value
\beq
\tau_{4;j} = \tau_{4,\rm av}= 4|abcd|\; .
\eeq

The four-qubit cluster state~\eqref{eq:2dcluster} is detected only by 
two independent four-qubit {\em SL}-invariants that vanish on product states.
Using the notation from Ref.~\cite{OS04}, these are 
\begin{eqnarray}\label{F42}
{\cal F}^{(4)}_2 &=&\left|{\mathfrak S}\left\{
                (\sigma_\mu\sigma_\nu\sigma_y\sigma_y)\bullet
                  (\sigma^\mu\sigma_y\sigma_\lambda\sigma_y)\bullet\right.
\right.
   \nonumber\\
   &&\left.\left.
   \qquad\qquad \bullet (\sigma_y\sigma^\nu\sigma_y\sigma_\tau) 
            \bullet (\sigma_y\sigma_y\sigma^\lambda\sigma^\tau)\right\}
	    \right| \\
{\cal F}^{(4)}_3 &=&\left |\bigfrac{1}{2}
                (\sigma_\mu\sigma_\nu\sigma_y\sigma_y)\bullet
                  (\sigma^\mu\sigma^\nu\sigma_y\sigma_y)
\bullet(\sigma_\rho\sigma_y\sigma_\tau\sigma_y) \bullet \right.
         \nonumber\\ &&\left.\  \bullet    
          (\sigma^\rho\sigma_y\sigma^\tau\sigma_y)
 \bullet(\sigma_y\sigma_\rho\sigma_\tau\sigma_y) \bullet
                (\sigma_y\sigma^\rho\sigma^\tau\sigma_y) \right|
 \ \ , \label{F43}
\end{eqnarray}
where $\mathfrak S$ indicates the symmetrization under four-qubit permutations.
It is interesting to note that the value of those measures exponentiated to 
homogeneous degree $4$ is $16 |abcd|/\sqrt{3}$ resp. $16 |abcd|$.
When we restrict ourselves to the family of telescope states from the third qubit,
we find
\nbeqa
{\cal F}^{(4)}_2&=& {\cal C}^{(4)}_{4;(1,3)}\left(\frac{7}{9} {\cal C}^{(4)}_{4;(1,3)}
                       +\frac{2}{9} {\cal C}^{(4)}_{4;(1,2)}\right) \\
{\cal F}^{(4)}_3 &=& \frac{1}{2} \left[{\cal C}^{(4)}_{4;(1,3)}\right]^2 
{\cal C}^{(4)}_{4;(1,2)}\; .
\neeqa

\section{Conclusions} 

We have analyzed possible extensions of the Coffman-Kundu-Wootters
monogamy equality to pure four-qubit states.
The known monogamy relations impose tight constraints 
on such extensions: the tripartite entanglement measure 
must coincide with the three-tangle on pure states, and 
the bipartite entanglement has to be measured by the 
two-tangle in order to respect the inequality due to Osborne and Verstraete.

We have presented
a detailed analysis of specific families of pure four-qubit states.
The example of the family (\ref{LOSU-states}) (as well as 
the state $\ket{\chi_1}$ in the Appendix) basically rules out
that a monogomy relation of the form (\ref{ourmonog}) can exist.
In particular, there are states that contain only 
four-tangle (vanishing two-tangle and three-tangle) while
the one-tangles are different. 
Since any reasonable four-tangle - as a global measure for entanglement -
should be permutation-invariant, this indicates clearly that  
a meaningful (i.e. state-independent) extension of the Coffman-Kundu-Wootters
monogamy relation to multipartite tangles does not exist. 
Even averaging over the one-tangles does not eliminate
this problem.

Nevertheless, there are interesting exceptions -- that is, families
of states which systematically do obey monogamy equalities. We have
called these states telescope states as their monogamy properties
can be retraced to those of the corresponding three-qubit states 
from which they 
can be generated by a qubit-doubling procedure. Interestingly,
the four-tangles in {\em these} states do coincide with the
values one can obtain from the known four-qubit polynomial 
invariants~\cite{OS04,DoOs08} which justifies to consider
them as four-tangles. We emphasize that the
four-tangle in general needs to be chosen according to the qubit-doubling
procedure applied to the three-qubit reference state.
Consequently, even for the four-qubit telescope states 
there is no {\em unique} extended monogamy relation of the form in
Eq.~(\ref{ourmonog}).

{\em Acknowledgments -- }
This work was supported by the German Research Foundation (SFB 631
and the Heisenberg program).
The authors would like to thank A.\ Uhlmann for stimulating discussions.

\section{Appendix}
Here we reconsider some pure four-qubit states previously
analyzed in Ref.~\cite{Bai07}. We begin our analysis with
\beqa\label{Baistate1}
\ket{\chi_1}&:=&\frac{1}{2}(\ket{0000}+\ket{1011}+\ket{1101}+\ket{1110})\\
\ket{\chi_2}&:=&a\ket{0000}+b\ket{0101}+c \ket{1000}+d\ket{1110}\label{Baistate2}\ \ .
\eeqa
The state $\ket{\chi_1}$ is symmetric under permutation of the last three qubits. 
In contrast to the conclusion of the authors of Ref.~\cite{Bai07},
this state has zero four-tangle, since every polynomial
$SL$ invariant gives zero for that state.
This can be easily checked by explicit evaluation of the
generating set of $SL$ invariants for four qubits~\cite{Luque02,Dokovic}. 
As observed in Ref.~\cite{Bai07} $\tau_3^{(234)}=0$,
since the reduced three-qubit density matrix is a mixture of 
a $W$ state with a product state.
For the other mixed three-tangles the reduced density matrix is a
rank-2 mixture of a {\em GHZ} state with a (biseparable) product state
such that the three-tangle can be computed by using the methods of
Ref.~\cite{EOSU}. We obtain
$\tau_3^{(123)}=\tau_3^{(124)}=\tau_3^{(134)}=1/4$.
Together with the one-tangles 
$\tau_1^{(1)}=3/4$, $  \tau_1^{(2)}= \tau_1^{(3)}= \tau_1^{(4)}= 1$ 
(the two-tangles vanish),
this leads to a valid monogamy relation for the first qubit only
whereas for qubits $2,3,4$ a mismatch of $1/2$ occurs.
It must be stressed at this point that no alternative
convex-roof extended function of the three-tangle can fix this
discrepancy. This is because the reduced density matrices
in this case are mixtures of $GHZ$ states and orthogonal
product states, and the corresponding characteristic curve is the convex function 
$p^2$ (here $p=1/2$) where $p$ is the weight 
of the {\em GHZ} state in the mixture.
In this particular case 
$\widehat{f(\tau_3)}\leq f(\widehat{\tau_3})$\footnote{In the case of non-convex characteristic 
curve, this inequality would apply to concave functions $f$.}, and then
$f^{\!-1}(\widehat{f(\tau_3)})\leq \hat{\tau}_3$.
This is a further proof that no monogamy relation of the form~\eqref{ourmonog}
including the three-tangle (in some form) can exist for pure states of 
more than three qubits.
In particular, this example indicates (in analogy to $\ket{{\Psi}_p}$ in Section III) 
that it must not be assumed that the residue $\cal R$ in the
monogamy relation be independent of the number of the distinguished qubit,
(as opposed to the approach in Refs.~\cite{Bai07,Bai08}). 

Finally we analyze $\ket{\chi_2}$ (cf.~Ref.~\cite{Bai07}).
This state has no four-tangle;
the three-tangles are calculated as~\cite{LOSU,KENNLINIE}
$\tau_3^{(123)}=4|ad|^2$, 
$\tau_3^{(124)}=4|bc|^2$, 
$\tau_3^{(134)}=4|bd|^2$, 
$\tau_3^{(234)}=0$,
the two-tangles are
$\tau_2^{12}=\tau_2^{13}=\tau_2^{14}=\tau_2^{34}=0$, $\tau_2^{23}=4|dc|^2$, 
 $\tau_2^{24}=4|ab|^2$
and the one-tangles are obtained as
$\tau_1^{(1)}=4(|bc|^2+|d|^2(|a|^2+|b|^2)$, 
$\tau_1^{(2)}=4(|a|^2+|c|^2)(|b|^2+|d|^2)$, 
$\tau_1^{(3)}=4|d|^2(1-|d|^2)$, 
$\tau_1^{(4)}=4|b|^2(1-|b|^2)$.
In this case, the monogamy relations are indeed fulfilled.
Since this state is at least not obviously a telescope state,
this might be a hint that also non-telescope states can satisfy
an extended monogamy relation.

\bibliography{gen-monogamy-arxiv}

\end{document}